\documentclass[]{spie}  
\usepackage[]{graphicx}
\usepackage{amsmath}

\pdfoutput=1
\usepackage[pdftex, a4paper, bookmarks=true, bookmarksnumbered=true, plainpages=false, breaklinks=true, colorlinks=true, urlcolor=blue, citecolor=blue, linkcolor=blue]{hyperref}

\title{GRAVITY: the Calibration Unit} 

\author{N. Blind\supit{a},  F. Eisenhauer\supit{a}, M. Haug\supit{a}, S. Gillessen\supit{a}, M. Lippa\supit{a}, L. Burtscher\supit{a}, O. Hans\supit{a}, F. Haussmann\supit{a}, S. Huber\supit{a}, A. Janssen\supit{a}, S. Kellner\supit{a}, Y. Kok\supit{a}, T. Ott\supit{a}, O. Pfuhl\supit{a}, E. Sturm\supit{a}, J. Weber\supit{a}, E. Wieprecht\supit{a}, A. Amorim\supit{b}, W. Brandner\supit{c}, G. Perrin\supit{d}, K. Perraut\supit{e}, C. Straubmeier\supit{f}
\skiplinehalf
\supit{a} Max-Planck-Institut f\"ur extraterrestrische Physik, 85748 Garching, Germany; \\
\supit{b} SIM, Fac. de Ci\^encias da Univ. de Lisboa, Campo Grande, Edif. C1, P-1749-016 Lisbon, Portugal;\\
\supit{c} Max-Planck-Institut f\"ur Astronomie, K\"onigstuhl 17, 69117 Heidelberg, Germany;\\
\supit{d} LESIA, Observ. de Paris Meudon, 5, place Jules Janssen, 92195 Meudon Cedex, France;\\
\supit{e} Institut de Plan\'etologie et d'Astrophysique de Grenoble (IPAG) UMR 5274, UJF-Grenoble 1/CNRS-INSU, Grenoble, France;\\
\supit{f} I. Physikalisches Institut, Universit\"at zu K\"oln, Z\"ulpicher Strasse 77, 50937 K\"oln, Germany.\\
}

\authorinfo{N. Blind: E-mail: blind@mpe.mpg.de}

\begin{document}
  \maketitle

\begin{abstract}
We present in this paper the design and characterisation of a new sub-system of the VLTI 2$^{nd}$ generation instrument GRAVITY: the Calibration Unit. The Calibration Unit provides all functions to test and calibrate the beam combiner instrument: it creates two artificial stars on four beams, and dispose of four delay lines with an internal metrology. It also includes artificial stars for the tip-tilt and pupil guiding systems, as well as four metrology pick-up diodes, for tests and calibration of the corresponding sub-systems. The calibration unit also hosts the reference targets to align GRAVITY to the VLTI, and the safety shutters to avoid the metrology light to propagate in the VLTI-lab. We present the results of the characterisation and validtion of these differrent sub-units. 
\end{abstract}



\section{INTRODUCTION}

GRAVITY is the second generation VLT Interferometer (VLTI) instrument for high-precision narrow-angle astrometry and phase-referenced interferometric imaging\cite{eisenhauer_2011a} . Its main goal is to observe highly relativistic motions of matter close to the event horizon of Sgr A*, the massive black hole at center of the Milky Way\cite{gillessen_2010a}. To do so, GRAVITY will combine four telescopes of the VLTI (either 4 ATs or 4 UTs), and will be assisted by four adaptive optics, plus a near-infrared fringe-tracker. During a typical observation, it will provide simultaneous interferometry of two objects within its field of view (2''  with UTs, 6" with ATs), and shall allow measurements of angles of order 10$\mu as$. We present here the design and characterisation of the Calibration Unit, a new sub-unit designed prior to the GRAVITY FDR. Its goal is to allow testing and validation of the different sub-systems behavior in the observatory conditions during integration in Garching, and, once in Paranal, to allow the daily calibration of the  beam combiner instrument. Since the initial design from the end of 2012, few change have been applied, the most important one concerning the optical injection unit.

\begin{figure}[t]
\centering
\includegraphics[width=.8\textwidth]{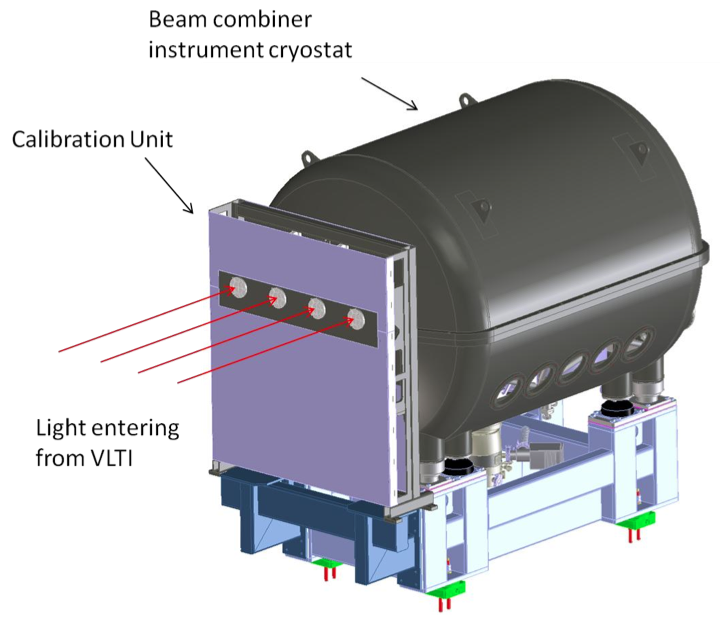}
\caption{Picture of the calibration unit in front of the GRAVITY. \label{fig:CUCryo}}
\end{figure}

The Calibration Unit is directly attached to the beam combiner instrument in front of the cryostat entrance window (Fig.~\ref{fig:CUCryo}). It should simulate:
\begin{itemize}
\item The broadband light from two stars for each of the four telescopes, that are used for routine test and calibration of GRAVITY during integration in Garching, and operations in Paranal;
\item A tip-tilt reference laser source for the tip-tilt guiding system\cite{pfuhl_2012a}, simmulating the source implemented in the VLTI star separators;
\item 4$\times$4 pupil tracking reference sources for the calibration of pupil tracking system\cite{amorim_2010a}, simulating the sources that will be finally implemented on the telescopes spiders in Paranal.
\end{itemize}
Its functions include four motorized delay lines, four metrology pick-up diodes (one per beam), and a coherent light source for monitoring the positions of the four delay lines. The calibration unit also provides the possibility to insert linear polarizing filter for calibration purpose, neutral densities, and rotating phase screens for simulating seeing. On top of these calibration and testing related functions, the calibration unit also hosts the safety shutters and the reference targets for the internal alignment of the instrument and, later, to the VLTI. Its optical specifications are summarised in Tab.~\ref{tab:spec}

\begin{table}[b]
\centering
\begin{tabular}{ll}
\hline
Operating wavelengths & 633\:nm, 1210\:nm, H+K-band\\
Separation of 2 stars & 1 to 2 arcsec \\
Strehl ratio in H-band & $>$ 90\%\\
Strehl ratio in K-band& $>$95\%\\
Extinction ratio of linear polarisers & $>1000$\\
Stroke of delay lines & $>$ 20\:mm\\
Incremental motion of delay lines & $<200$\:nm\\
\hline
&\\
\end{tabular}
\caption{Calibration Unit specifications.}\label{tab:spec}
\end{table}

\section{MECHANICS}

\begin{figure}
\centering
\includegraphics[width=1.\textwidth]{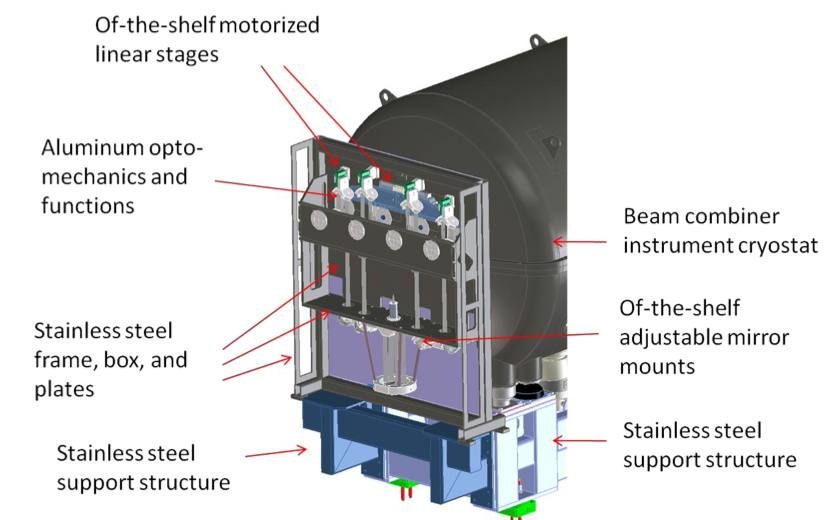}
\caption{Overview of the calibration unit mechanics.}\label{fig:CU_mecha}
\end{figure}

An overview of the calibration unit mechanics is presented on Fig.~\ref{fig:CU_mecha}.
The structural elements of the calibration unit are made from stainless steel. These structural elements are the support structure, the frame structure, and the mounting plates.
The optomechanics and functions are mostly made from aluminum.

A first linear stage allows removing/entering fold-mirrors to feed the beam combiner instrument with starlight or with the light from the calibration unit. A second linear stage can insert linear polarization filters and phase screens in the beam. A set of four small linear stages move the delay line mirrors. All moving functions are of-the-shelf commercial products.

Two mirrors per beam are mounted on commercial of-the-shelf mirror mounts to align each collimated beam. All mirrors are hold using spring loaded optical mounts. The mounts of the largest mirrors (1x 150\:mm launch telescope parabola, 2x  50x80 mm, and 2x 30x50\:mm flat mirrors) are based on a three point support. The smallest 50\:mm diameter flat mirrors use a 0.3\:mm kapton foil to avoid stress induced deformations.

\section{OPTICAL LAYOUT}
\label{part:optical_layout}

\begin{figure}
\centering
\includegraphics[width=.85\textwidth]{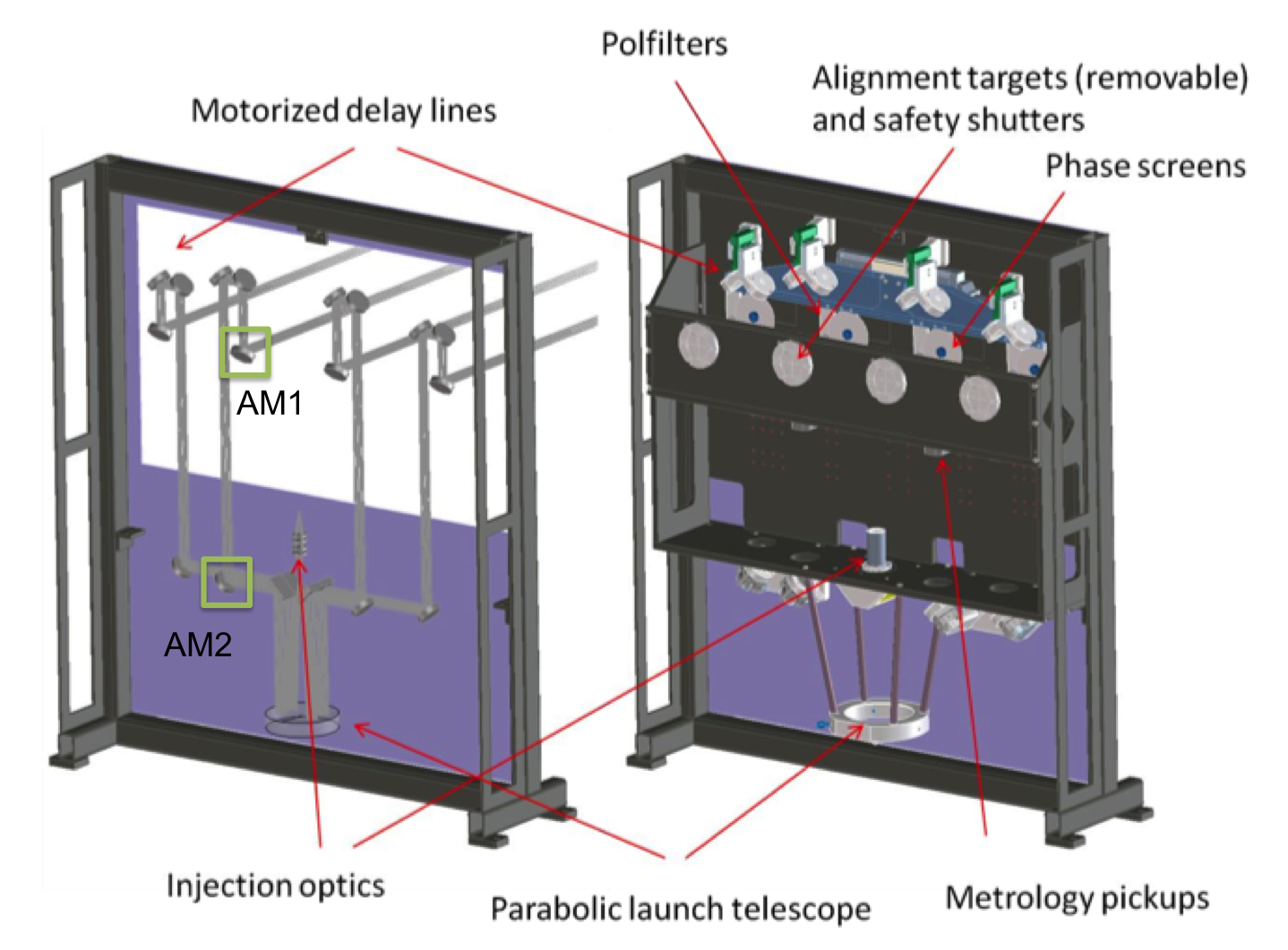}\vfill
\includegraphics[trim=0 5mm 0 9cm, clip=true, width=1.\textwidth]{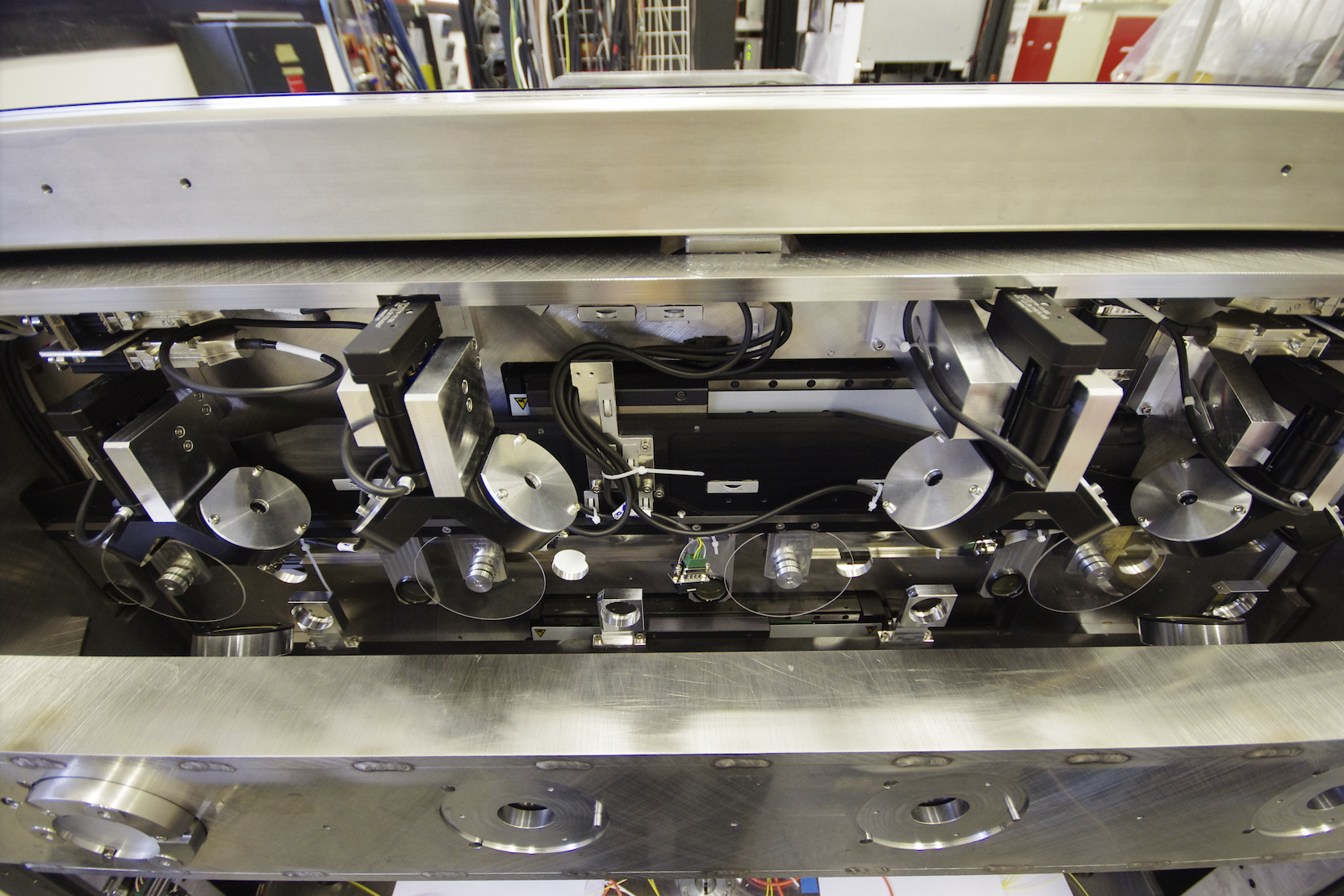}
\caption{{\it Top} -- Schematics of the Calibration Unit optical path on the left (Sect.~\ref{part:optical_layout}), and functions on the right (Sect.~\ref{part:functions}). AM1 and AM2 (green squares) are the mirrors used for the Calibration Unit alignment (see Sect~\ref{part:alignment}), here presented for the beam \#2. {\it Bottom} -- Picture of the Calibration Unit top part where are placed the four delay lines, the Mode Selector stage and its four phase screens and four polarisers. The bottom part of the picture shows the GRAVITY reference plate, with, on the left, one of the target mirror mounted for internal alignment.\label{fig:CUIn}}
\end{figure}

Fig.~\ref{fig:CUIn} presents an overview of the calibration unit optical layout, which consists of:
\begin{itemize}
\item a single, uncoated lens injection optics;
\item a bare gold coated launch mirror providing four collimated beams to GRAVITY;
\item a set of 5 flat gold coated mirrors, relaying the launch telescope collimated beams to GRAVITY.
\end{itemize}
%

The optical parameters of the calibration unit are summarized by:
\begin{itemize}
\item	Collimated beam / exit pupil diameter: 18 mm;
\item	Beam separation: 240 mm;
\item	Focal length of launch telescope: 300 mm;
\item	Magnification of injection optics: 1/2.5;
\item	Two point sources separated by 1.5 arcsec on sky for UTs, i.e. by $\sim$2\:mm in the launch telescope focal plane;
\item	Pupil stop in injection optics is 49.4 mm in front of launch telescope focus;
\item	Exit pupil (i.e. fiber coupler TT mirror) is situated 739.2 mm behind the cryostat entrance window.
\end{itemize}

\begin{figure}
\centering
\includegraphics[width = 0.5\textwidth, trim=9cm 0 7cm 2.8cm , clip=true]{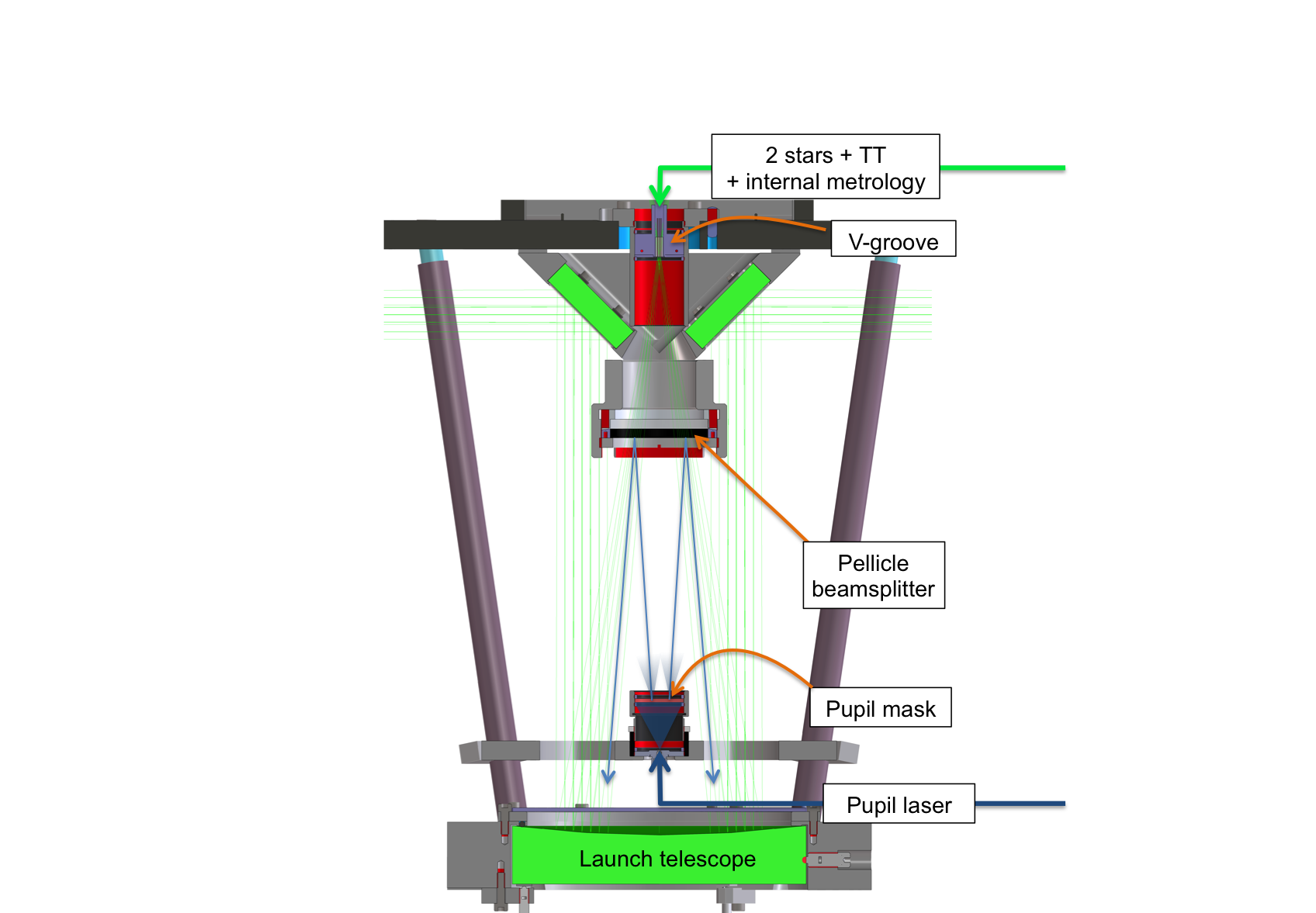}
\caption{Schematics of the calibration unit injection stage. On top is the 3-fiber V-groove, fed with continuum light and internal metrology on the 2 external fibers (the stars), and tip-tilt visible laser on the central, on-axis fiber. On-axis rays are represented in green from a Zemax file for a direct injection scheme. The pupil tracker source is placed on top of the launche telescope: it illuminates in direction of the V-groove, before being redirected to the launch telescope and GRAVITY by a pellicle beamsplitter. Light path is represented in blue.}\label{fig:CU_injection}
\end{figure}

All light sources are launched from the injection unit (Fig.~\ref{fig:CU_injection}).

\subsection{Launch telescope and beam relay}
\label{part:beam_relay}


The launch telescope is an f\#2 parabolic mirror with 300\:mm focal length, with an RMS wavefront error is less than 15\:nm. The calibration optics uses only four $\sim$30\:mm diameter parts of the collimated beam. It is illuminated by two single mode fibers, separated by 2\:mm, corresponding to two stars separated by 1.5" for UTs.

The beam relay is made of five flat, gold protected mirrors for each beam. Each mirror has an optimal optical flatness of $\lambda/20$ (@$\lambda=$633\:nm). The optical path lengths of all four arms are identical.

\subsection{Injection optics}

\begin{figure}
\centering
\includegraphics[trim=1cm 3cm 2cm 3cm, clip=true, width=0.7\textwidth]{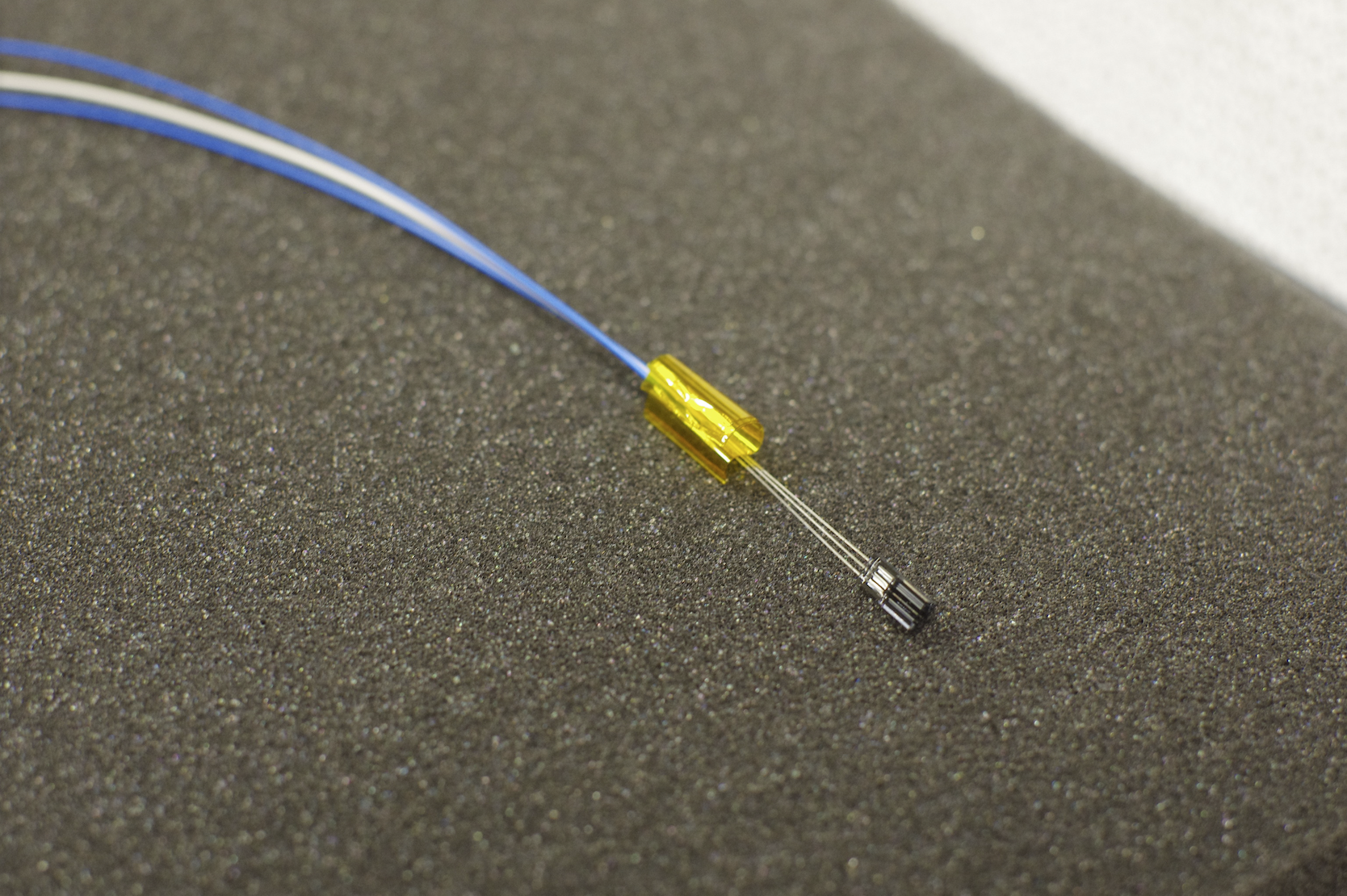}
\caption{Picture of the final V-groove injecting the 2 stars (external fibers, 2\:mm away from each other), and the tip-tilt reference (central multimode fiber) in the calibration unit. \label{fig:vgroove}}
\end{figure}

The injection optics consists first of a V-groove from OZ Optics, containing 3 fibers with a 1\:mm pitch (Fig~\ref{fig:vgroove}):
\begin{itemize}
\item The two external single-mode fibers simulate the calibration stars distant by 1.5" on sky (UT);
\item The central one is a multimode fiber from which is injected the tip-tilt reference light.
\end{itemize}
The two single-mode fibers have a numerical aperture of only 0.22, insufficient to uniformly illuminate the launch telescope. The injection optics includes a plano-concave CaF$_2$ lens from Thorlabs ($f=-18$\:mm) demagnifying the fiber field by a factor 2.5. The illumination gradient in the 30\:mm beam is then reduced from 3 (for a direct fiber injection) to 1.3 Peak-to-Valley (P-V). This has the additional advantage to pick up more light from the gaussian beam at the launch telescope subapertures.

\subsection{Pupil tracker sources}

\begin{figure}[t]
\begin{minipage}[c]{0.55\textwidth}
\centering
\includegraphics[width=.95\textwidth]{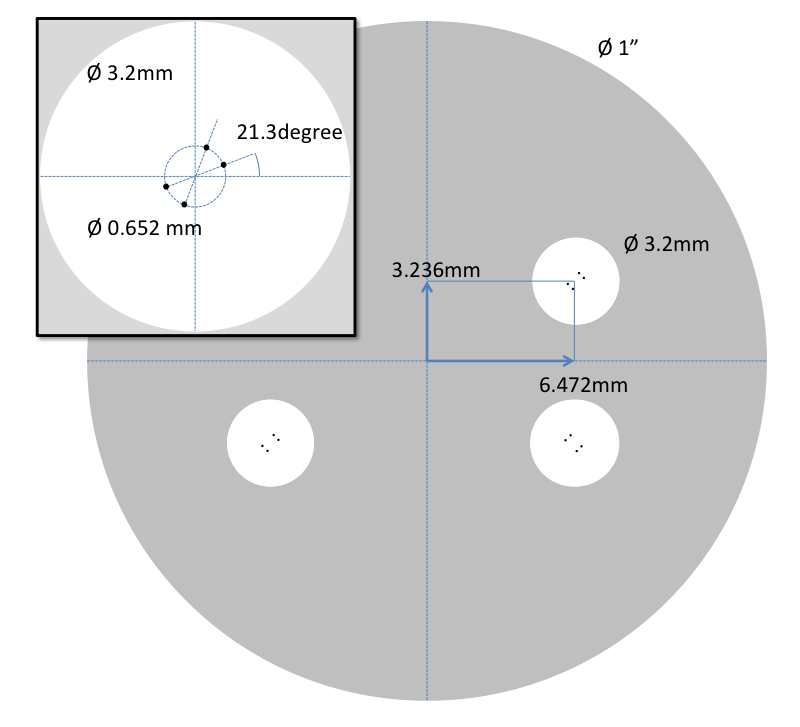}
\caption{
Drawing of the 1" mask (grey area), with the position of the telescopes represented by the white area. The top left part zooms on one of them, the laser pierced holes simulating the pupil trackers sources being at the position of the black dots. \label{fig:CUPupil}}
\end{minipage}
\hfill
\begin{minipage}[c]{0.4\textwidth}
\centering
\includegraphics[width=.85\textwidth]{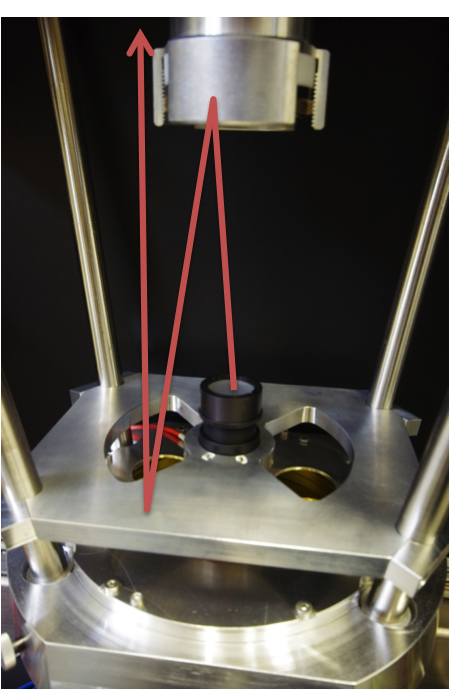}
\caption{Picture of the pupil tracker source injection stage. The beamsplitter reflects 10\% of the pupil light at 1210\:nm, while the stars and tip-tilt reference sources launched by the injection unit goes through it. All the beams are then collected by the launch telescope and sent to GRAVITY.\label{fig:CUpupil_pic}}
\end{minipage}
\end{figure}

The calibration unit must provide artificial reference sources for the Acquisition Camera pupil tracker\cite{amorim_2010a} . The source consists of a multimode fiber illuminating a laser pierced aluminium mask: four series of four holes are made in this mask, corresponding to the actual position of the pupil tracker sources in the UTs spiders (Fig.~\ref{fig:CUPupil} and \ref{fig:CUpupil_pic}). This simple optics is positioned on top of the launch telescope: it illuminates towards the injection optics, and is reflected by a pellicle beamsplitter to the launch telescope. The position of the mask corresponds to the launch telescope pupil, with respect to the pellicle beamsplitter.


%

\subsection{Optical quality}

The nominal image quality of the calibration unit optics is $>95\%$ over the whole K-band, and all across the 2" field of view. Therefore the final image quality will be fully dominated by manufacturing errors.

\subsubsection{Tolerances}

The nominal wavefront error is less than 15 nm RMS on the launch telescope. The specification of the surface accuracy across the clear aperture of the parabolic mirror is $< \lambda/10$ P-V @ 633 nm. Therefore the wavefront error of the reflected beam (factor 2) is better than 126 nm P-V. Under the conservative assumption that the wavefront error is dominated by astigmatism and spherical aberrations, the RMS wavefront error is about a factor 4 smaller than the P-V, i.e. the wavefront error from the launch telescope reflected beam is $\sim$31 nm RMS. Assuming that the nominal wavefront errors and manufacturing errors are independent and add in square, the total wavefront error from the launch telescope is $\sim 34$\:nm RMS.

The beam relay consists of five flat mirrors that do not introduce wavefront errors in a collimated beam. Therefore the wavefront error is fully dominated by manufacturing errors. 
Each beam relay consists of five flat mirrors with a surface accuracy specification of $\lambda/20$ P-V  @ 633 nm, or 32\:nm. Following the same arguments as for the launch telescope parabola (factor 2 from reflection, factor $1/4$ from P-V to RMS) the wavefront error introduced per mirror is 16\:nm. Assuming that the wavefront errors from the five mirrors are independent and add in square, the wavefront error of the beam relay optics is $\sim$35 nm RMS.

The surface quality of the lens in the injection optics is specified at $\lambda/2$ at 633\:nm. Because the actual beam diameter for each of the four telescopes is only $\sim$\:1/8 of the lens diameter, the optical quality can be estimated to at least $\lambda/5$ or $\lambda/10$ for the useful sub-apertures.

The stellar beams are also going through the pellicle beamsplitter used for the pupil tracker sources.  Given its thickness (only few microns) it does not introduce additional wavefront error.

Assuming that the wavefront errors from the above sub-units are independent, the total wavefront error of the calibration unit optics is:
\begin{itemize}
	\item For a direct injection with a fiber at the launch telescope focus, the total wavefront error from launch telescope and beam relay is $\sim$\:48\:nm, i.e. a Strehl ratio $>$\:95\% all across the 1.4 - 2.5\:$\mu$m wavelength range.
	\item Including the injection optics, the total wavefront error is now $\sim 90$ to 140\:nm, i.e. a Strehl ratio of $>$\:85\% at wavelengths 1.4 - 1.8\:$\mu$m, and still $>$\:95\% at wavelengths 1.95 - 2.45 $\mu$m.
\end{itemize}

Zemax simulations also shows good performance of the single lens injection unit design, with theoretical Strehl ratio of 80\% at 633\:nm, and $> 95\%$ above 1\:$\mu$m. The lateral dispersion across H and K bands is also well contained within the diffraction spot.


\subsubsection{Characterisation}

We measured the optical quality of the two stars and four beams of the Calibration Unit with a Shack-Hartmann wavefront sensor and a visible laser at $\lambda = 633\:$nm. The measurements include the injection optics, the launch telescope and two of the flat mirrors. The remaining three mirrors were not accessible at the time of the measurements, so that we add a flat relay mirror to access the beams. Once tilt and defocus errors removed, we measured a wavefront error contained in the range of 40 to 50\: nm for the two stars and the four telescope. Removing the relay mirror to the wavefront sensor ($\lambda/10$ optical flatness) and adding the 3 missing mirrors ($\lambda/20$ flatness), the total wavefront error of the calibration unit rises to 60 to 70\:nm, from which we estimate a Strehl ratio well above 90\% for all beams in the H and K bands (Tab.~\ref{tab:optical_quality}). Under the pessimistic assumption of an optical flatness of $\lambda/10$ for the 3 relay mirrors, performances are still in specification, with Strehl ratio $\geq$90\% in H and K bands.  We observed slightly better performance for the SCI star, probably because of decentering of the injection unit lens with respect to the V-groove.

\begin{table}
\centering
\begin{tabular}{lcccc}
\hline \hline
$\lambda$ [$\mu$m] & 0.633            & 1.5                & 2.0                     & 2.5                \\
\hline
Strehl ratio                 & 65$\pm$7\% & 93$\pm$2\% & 96$\pm$1\%  & 97$\pm$1\% \\
\hline
\end{tabular}
\caption{Average measured Strehl ratio for the 2 off axis stars, with error given as min and max dispersion among the 8 samples (4 telescopes and 2 stars).} \label{tab:optical_quality}
\end{table}

\subsection{Polarisation behavior}

The optical design of the four beams is fully symmetric in order to limit differential birefringence of the optical train that could severely impact the fringe contrast during calibrations. Measurements of the output polarisation state while injected linearly polarised laser showed a differential retardance below 5$^\circ$, i.e. a loss of less than 1\% in fringe contrast. The initial set of mirrors used for these measurements were of bare gold coated. They recently showed an important aging, and will be soon changed for gold protected ones. These measurements will then be repeated.

\subsection{Alignment}
\label{part:alignment}

The Calibration Unit is aligned against the target mirror placed at its entrance. It disposes of two adjustable mirrors for pupil and field alignments, the three remaining flat mirrors being in fixed position. As seen from GRAVITY, the first one (on the Beam Selector stage, AM1 on Fig.~\ref{fig:CUIn}) is centered against the metrology pick-ups. The second one (AM2 on Fig.~\ref{fig:CUIn}) is then centered on the central fiber of the injection unit V-groove. As these two mirrors are not placed in conjugated pupil and field planes, pupil and field alignment are not independent. Between them are also placed the delay lines, which can lead to a different pupil offset of the four beams. The pupil offset is little (less than 1\:mm), and therefore does not vignet the beam in the fiber coupler or acquisition camera. However the artificial pupil tracker sources are not perfectly aligned against each others on the acquisition camera. A redesign of the mask with re-optimized pupil tracker sources positions will solve the problem.

\section{LIGHT SOURCES}

%
%

All coherent sources will be installed in one of the electronic cabinet in the combined Coude laboratory. The two continuum sources are installed directly in the calibration unit. An overview of the light sources and beam transport is shown in Fig~\ref{fig:CU_injection}:
\begin{itemize}
\item {\bf Continuum} -- The two continuum light sources are Quartz-Tungsten Halogen lamp from B\&W Tek (BPS101) with a color temperature of 2800 K. Depending on the applied voltage, they generate a star of magnitude up to K$_{max}^{FT}$ = 2.5 in the Fringe Tracker channel when considering a VLTI transmission of 28\% and an average Strehl ratio of 30\% (Fig.~\ref{fig:starmag}). The Science channel includes an additional coupler to provide the "star" light and internal metrology via the same fiber: with a transmission of only 10\%, it decreases the star luminosity by 2.5 magnitudes, i.e. K$_{max}^{SCI}$ = 5.0.
\begin{figure}
\begin{minipage}[t]{0.49\textwidth}
\centering
\includegraphics[width=0.9\textwidth]{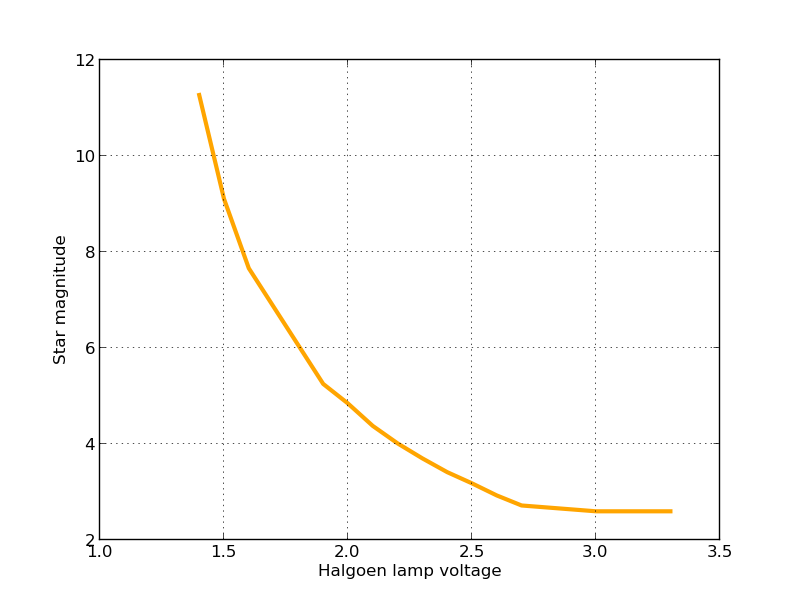}
\caption{
Star magnitude as function of the halogen lamp voltage for the Fringe Tracker star. The additional coupler on the Science star increases the magnitude by $\delta$K = +2.5. }\label{fig:starmag}
\end{minipage}
%
%
\begin{minipage}[t]{0.49\textwidth}
\centering
\includegraphics[width=0.9\textwidth]{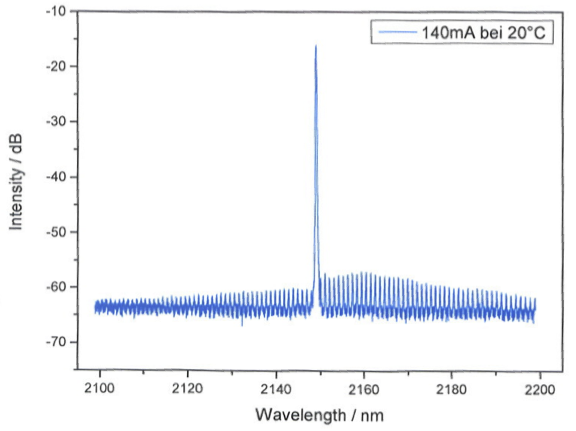}
\caption{Sacher laser calibrated spectrum.}\label{fig:sacherlaser}
\end{minipage}
\end{figure}
\item {\bf Pupil guiding light source} -- The pupil guiding light source of the calibration unit is a CW single mode injection semiconductor laser at $\lambda = 1210$\:nm from Frankfurt Laser Company (FTLD-1200-05S), delivering an optical power of $\sim$4\:mW. It is identical to the pupil guiding light source installed at the telescopes.
\item {\bf Tip/Tilt guiding light source} -- The tip/tilt guiding light source of the calibration unit is a single-mode pigtailed laser diode from Thorlabs (LPS 660-FC). It is similar to the tip/tilt guiding light source installed at the star separators, but with reduced power (7.5 mW instead of 60 mW).
\item {\bf Internal metrology} -- In order to calibrate the spectral dispersion of the Fiber Differential Delay Lines, it is necessary to monitor the calibration unit delay lines position when doing measurements on the Science spectrometer. It is done thanks to a Distributed FeedBack laser diode at $\lambda = 2150\:$nm from SacherLaser (Fig.~\ref{fig:sacherlaser}), whose wavelength can be tuned in a $\pm$2\:nm range. The wavelength stability is better than 1\:nm thanks to a high precision electronics delivering a constant current, and an internal temperature controller.
\end{itemize}

Light is transported from the different sources with standard silica fibers (Thorlabs PM2000) ensuring a transmission higher than 80\% up to 2200\:nm. The coupler used to combine the internal metrology to the continuum sources on the Science fiber, is a 1-to-2 O/E-land WideBand Splitter, based on an internal infra-red 50/50 dichroic from Edmund Optics, and connectorized on the three ends to Thorlabs PM2000 fibers and collimators. Their transmission has been estimated to $\sim$10\% (from input to 1 output). As silica fibers absorption increases dramatically beyond 2200\:nm ($\sim$\:3\:dB/m at 2500\:nm), we limited their total length to $2$\:m, so that the calibration stars are not  excessively blue.

\section{FUNCTIONS}
\label{part:functions}

A schematics of the different functins provided by the calibration unit are visible on Fig.~\ref{fig:CUIn}.

\subsection{Beam selector}

The beam selector is a motorized linear translation stage which moves a mirror in and out each beam of the Calibration Unit to feed the beam combiner instrument with either the light from the Calibration Unit or of the VLTI respectively. A third position allows putting neutral density filters between the VLTI and GRAVITY for observing bright stars like Betelgeuse.

\subsection{Mode selector}

The mode selector is a motorized linear translation stage which can optionally move a linear polarization filter or a rotating phase screen into the beams:
\begin{itemize}
\item {\bf Phase screens} -- The rotating phase screens simulate the high-order wavefront residuals after the VLTI AO correction. This function is not required for the instrument calibration at the observatory, but only for system level tests.

The phase screens used in GRAVITY are those from the PRIMA test bed\cite{sahlmann_2007a}. They have encoded on their surface a bi-dimensional aberration with a spatial distribution representing the residual atmospheric wavefront aberrations after correction by the Adaptive Optics system of the VLTI while observing a K=9 magnitude star (Fig.~\ref{fig:ps_CU}). Two versions of the phase screens are available, one simulating the AO residuals for seeing conditions of 0.65 arcsec and a coherence time of 4\:ms, and the second one simulating the AO residuals for seeing conditions of 1 arcsec and a coherence time of 2\:ms.
\begin{figure}
\centering
\hfill
\includegraphics[width=0.4\textwidth]{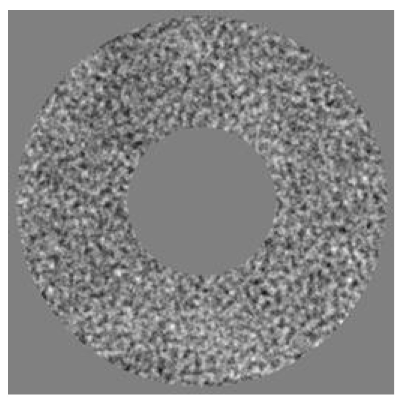}\hfill
\includegraphics[width=0.4\textwidth]{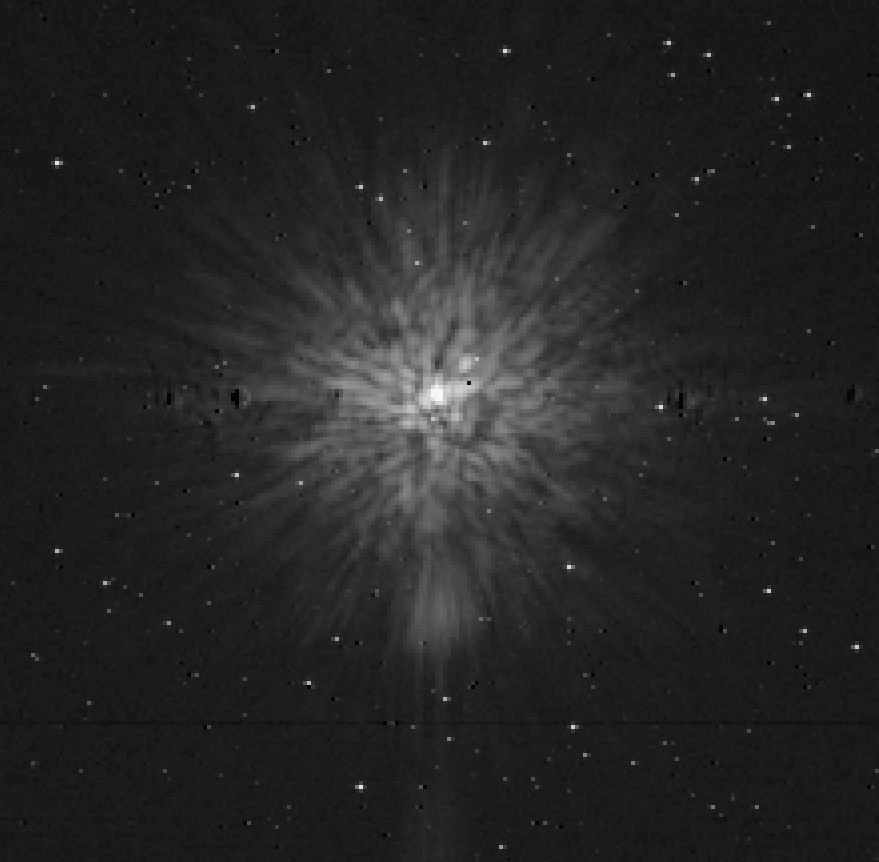}
\hfill
\caption{{\it Left:} Wavefront map of the phase screen. The internal and external radii of the screen are 15 and 45\:mm respectively. {\it Right:} Acquisition camera image of a calibration unit star with the phase screen in the path (log scale).}\label{fig:ps_CU}
\end{figure}

\item {\bf Polarisers} -- The calibration unit can optionally move linear polarization filters from Codixx, into the 4 beams, aligned to the vertical axis of the instrument. They are guaranteed for an extinction ratio of 1:1000 from 500 to 1500\:nm, and $\sim$\:1:10000 in the K-band.
\end{itemize}

\subsection{Safety shutters} 
Safety shutters are installed at the entrance of the calibration unit for blocking the metrology light when GRAVITY is not observing. They will be connected to the interlock system of the VLTI.

\subsection{Alignment targets} 
Target mirrors can be installed in front of the safety shutters and are the alignment reference of GRAVITY. We use four mirror targets from Brunson to align GRAVITY and the calibration unit to each other. Once in Paranal, these mirrors will be mounted the other way round to finally align GRAVITY  to the VLTI.

\subsection{Metrology pick-up}
\begin{figure}
\centering
\includegraphics[trim=1cm 3cm 2cm 5cm, clip=true, width=0.4\textwidth]{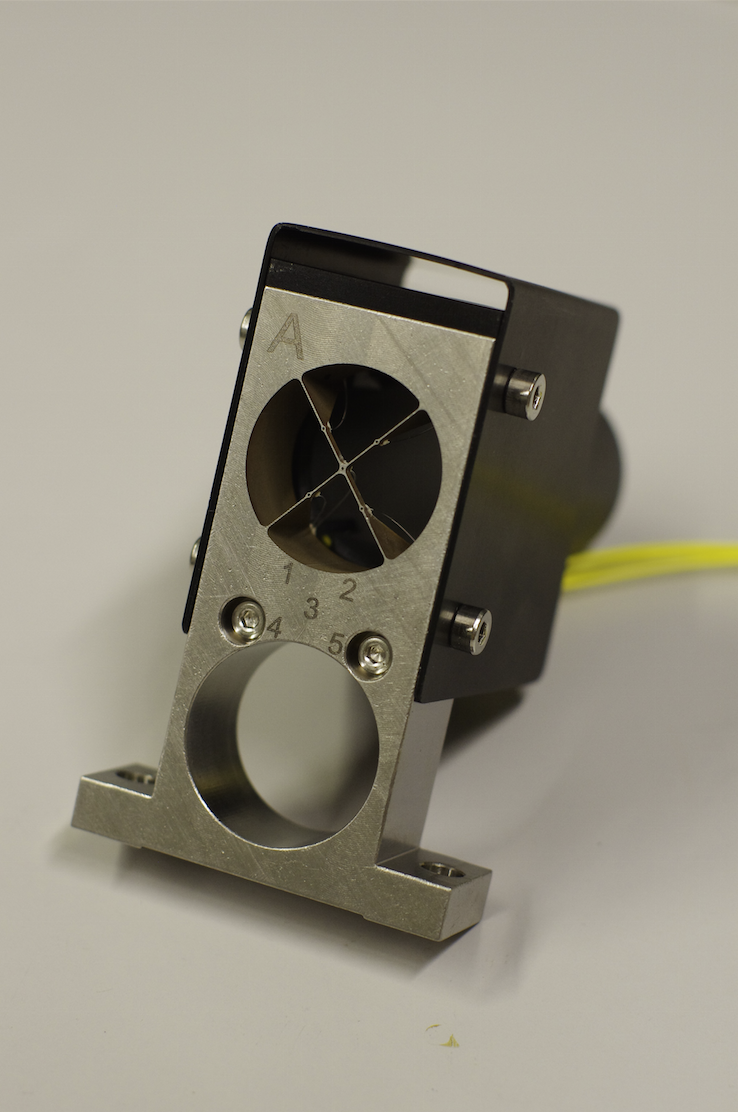}
\caption{Picture of one prototype metrology pick-up. Five fibers are glued into the spider: the central one is used as metrology pick-up, and is directly connected to a metrology receiver box in the calibration unit chassis. The four external ones were initially simulating each of the pupil tracker sources, but are not part of the final pick-up. These prototype pick-ups are fully functionnal and will be used as spare parts.}\label{fig:met_pickup}
\end{figure}

 The calibration unit is equipped with four metrology pickup fibers located in the middle of each beam (Fig.~\ref{fig:met_pickup}). The calibration unit metrology receivers simulate the metrology feedback from the telescopes. The fiber is a standard graded index multi-mode fiber, with core diameter of 62.5\:$\mu$m, corresponding to 1/288 of the pupil diameter. This is $<$\:1/7 of the smallest metrology fringe spacing (the pupil contains up to 40 fringes): the according contrast loss of the metrology signal is thus negligible. The fibers send the light to four replica of the metrology receiver boxes\cite{gillessen_2010a} mounted inside the calibration unit. The subsequent signal path is identical to the UT metrology system. 

\subsection{Delay lines}
For interferometric calibration purposes, the calibration unit is also equipped of one delay line per telescope. They consist of a roof mirror (made of two flat mirrors) mounted on a translation stepper motor. The motor stroke is 14\:mm, i.e. an optical path modulation of 28\:mm each. Their linearity was measured to be better than 0.1\% over the full stroke.

\section{CONCLUSION}

The Calibration Unit was designed to provide test and calibration functions to the beam combiner instrument. It was mounted in front of the cryostat in March 2013, and since then, was thoroughly tested and partly redesigned to reach the required performances, as demonstrated in this paper. It is now used for the routine operations and tests of GRAVITY during its integration at the MPE, and will be used as such till its shipment in Paranal. Once there, it will be used for the final alignment of GRAVITY to the VLTI, and on a daily basis for all the photometric and interferometric calibration procedures.

\bibliographystyle{spiebib}
\bibliography{calibration_unit}

\end{document}